\documentstyle[twoside,fleqn,espcrc2,epsfig]{article}

\def\gg{\gamma\gamma}
\def\sigmagg{\sigma_{\gg}}
\def\ee{\mbox{${\rm e}^+{\rm e}^-$}}

\def\ssgg{s_{\gamma\gamma}}
\def\sqee{\mbox{$\sqrt{s}_{\rm ee}$}}
\def\Lgg{\mbox{$L_{\rm \gamma\gamma}$}}
\def\Wgg{\mbox{$W$}}

\def\Wvis{\mbox{$W_{\rm vis}$}}

\def\ee{\mbox{${\rm e}^+{\rm e}^-$}}

\def\WECAL{W_{\rm ECAL}}

\def\thetamax{\theta_{\rm max}}

\def\Zzero{\ifmmode {{\mathrm Z}^0} \else {${\mathrm Z}^0$} \fi}
\def\ppbar{\overline{\mbox p}\mbox{p}}

\def\pz{\phantom{0}}
\def\pzz{\phantom{00}}

\def\dW{{\rm d}\sigma _{\rm ee}/{\rm d}W}
\def\dsee{{\rm d}\sigma _{\rm ee}}

\def\ggh{\mbox{${\rm \gamma \gamma \rightarrow hadrons}$}}

\newcommand{\AmS}{{\protect\the\textfont2
  A\kern-.1667em\lower.5ex\hbox{M}\kern-.125emS}}

\title{
Total Hadronic Cross-Section for Photon-Photon Interactions at 
LEP \footnotemark[1]
}

\author{Frank W\"ackerle\address{Fakult\"at f\"ur Physik, 
Albert-Ludwigs-Universit\"at Freiburg, Germany}}

\begin{document}

\begin{abstract}
The total hadronic cross-section $\sigmagg$ for 
the interaction of real photons, $\gg\rightarrow\mbox{~hadrons}$, 
is extracted from a measurement of the cross-section of the 
process $\ee\rightarrow\ee\gamma^*\gamma^*\rightarrow(\ee+\mbox{~hadrons})$
using a luminosity function for the photon flux and form factors 
for extrapolating to $Q^2=0$. 
The data was taken with the OPAL detector at LEP at $\ee$
centre-of-mass energies $\sqee=161$~GeV and $172$~GeV.
In the energy range $10 \le W \le 110$~GeV the
total hadronic $\gg$ cross-section $\sigmagg$ is consistent
with the Regge behaviour of the total cross-section observed
in $\gamma$p and hadron-hadron interactions.
\end{abstract}

\maketitle

\section{Introduction}
At high $\gg$ centre-of-mass energies $W=\sqrt{s}_{\gg}$ 
the total  cross-section for the production of hadrons in the interaction of 
two real photons is expected to be dominated by interactions
where the photon has fluctuated into an hadronic state. 
Measuring the $\sqrt{s}_{\gg}$ dependence of the total hadronic $\gg$ cross-section
$\sigmagg$
should therefore improve our understanding of the hadronic nature of
the photon and the universal high energy behaviour of total 
hadronic cross-sections.

Before LEP the total hadronic $\gg$ cross-section has only been measured
for $\gg$ centre-of-mass energies $W$ below 10~GeV by
PLUTO~\cite{bib-pluto}, TPC/2$\gamma$~\cite{bib-tpc} and
the MD1 experiment~\cite{bib-md1} where the high energy
rise of the total cross-section could not have been observed.
Using LEP data taken at $\ee$ centre-of-mass energies $\sqee=130-161$~GeV,
L3~\cite{bib-l3} has demonstrated that
the total hadronic $\gg$ cross-section in the range 
$5 \le W \le 75$~GeV 
is consistent with the universal Regge behaviour of total cross-sections.
We present a measurement of the total hadronic $\gg$ 
cross-section in the range $10<W<110$~GeV 
using OPAL data taken at $\sqee=161$~GeV and $172$~GeV. 
\footnotetext[1]{Talk given at the
XXVII INTERNATIONAL SYMPOSIUM ON MULTIPARTICLE DYNAMICS,  
Laboratori Nazionali di Frascati - INFN,
Frascati (Rome), Italy, 8-12 September 1997} 
\section{Kinematics}
\label{sec-kine}
The kinematics of the process $\ee\rightarrow(\ee+\mbox{hadrons})$
at a given \sqee{} can be described by the negative square of the 
four-momentum transfers, $Q_i^2=-q_i^2$, carried by the two ($i=1,2$) 
incoming photons and by the square of the invariant mass of the hadronic final 
state, $W^2=\ssgg=(q_1+q_2)^2$~\cite{bib-kolanoski}. 
Events with detected scattered electrons (single-tagged or
double-tagged events) are excluded from
the analysis. This anti-tagging condition defines an effective upper
limit $Q^2_{\rm max}$ on the values of $Q_{i}^2$ for both photons. 
This condition is met when the scattering angle $\theta'$ of
the electron is less than the angle $\thetamax=32$~mrad between the beam axis 
and the inner edge of the acceptance of the detector or if the energy of
the scattered electron is smaller than the minimum energy of 35~GeV
required for the tagged electron.

\section{Monte Carlo simulation}
\label{sec-MC}
The leading order (LO) QCD 
Monte Carlo generators PYTHIA 5.722~\cite{bib-pythia}
and PHOJET 1.05c \cite{bib-phojet} are used to simulate
photon-photon interactions. PYTHIA is based on a model
by Schuler and Sj\"ostrand~\cite{bib-GSTSZP73}
and PHOJET has been developed by Engel based on
the Dual Parton model (DPM)~\cite{bib-dpm}.
The SaS-1D parametrisation of the parton distribution 
functions~\cite{bib-sas} is used 
in PYTHIA and the leading order GRV parametrisation \cite{bib-grv}
in PHOJET. 
The fragmentation and decay of the parton final state is handled 
in both generators by the routines of JETSET 7.408 \cite{bib-pythia}.

\section{Event selection}
Two-photon events are selected by requiring that
the visible invariant hadronic mass, $\WECAL$, 
measured in the electromagnetic calorimeter (ECAL), 
has to be greater than 3 GeV. At least 3 tracks must have been found
in the event and the sum of all energy deposits in the ECAL and the 
hadronic calorimeter (HCAL)
has to be less than 45 GeV.
The missing transverse energy of the event
measured in the ECAL and the forward
calorimeters (FD) has to be less than 5 GeV.
No track in the event has a momentum greater than 30 GeV/$c$.
Finally, to remove events with scattered electrons in the FD or the
silicon tungsten calorimeter (SW),
the 
energy 
measured in the FD has to be less than
50 GeV and the 
energy
measured in the SW
less than 35 GeV (anti-tagging condition).

Additional cuts are applied to reject beam-gas and beam-wall background.
On average the trigger efficiency for the lowest $W$ range, $10<W<30$~GeV,
is greater than 97\% and it approaches 100\% for larger values of $W$.

We use data corresponding to an integrated luminosity of
9.9 pb$^{-1}$ at $\sqee=161$~GeV and 10.0 pb$^{-1}$ at $\sqee=172$~GeV.
After applying all preselection cuts $55169$ events remain.
From the Monte Carlo (MC) it is estimated that after all cuts 
about 4\% of all remaining events are e$\gamma$ processes with
$Q^2>1$~GeV$^2$.
The background from other processes apart from
beam-gas and beam-wall interactions
amounts to less than 1\%.

\section{\boldmath $W$ \unboldmath reconstruction}
For measuring the total hadronic $\gg$ cross-section $\sigmagg$
the value of $W$ must be reconstructed from the hadronic final state.
After the event selection a matching algorithm is applied
in order to avoid double counting of particle momenta. 
The matching algorithm uses all the information of the 
ECAL, the HCAL, the FD and the SW
calorimeters, as well as the tracking system.
The four-momenta of the detected particles are used to calculate the visible
invariant mass \Wvis{}.

\begin{figure}[htb]
   \begin{center}
      \mbox{
          \epsfxsize=0.465\textwidth
          \epsffile{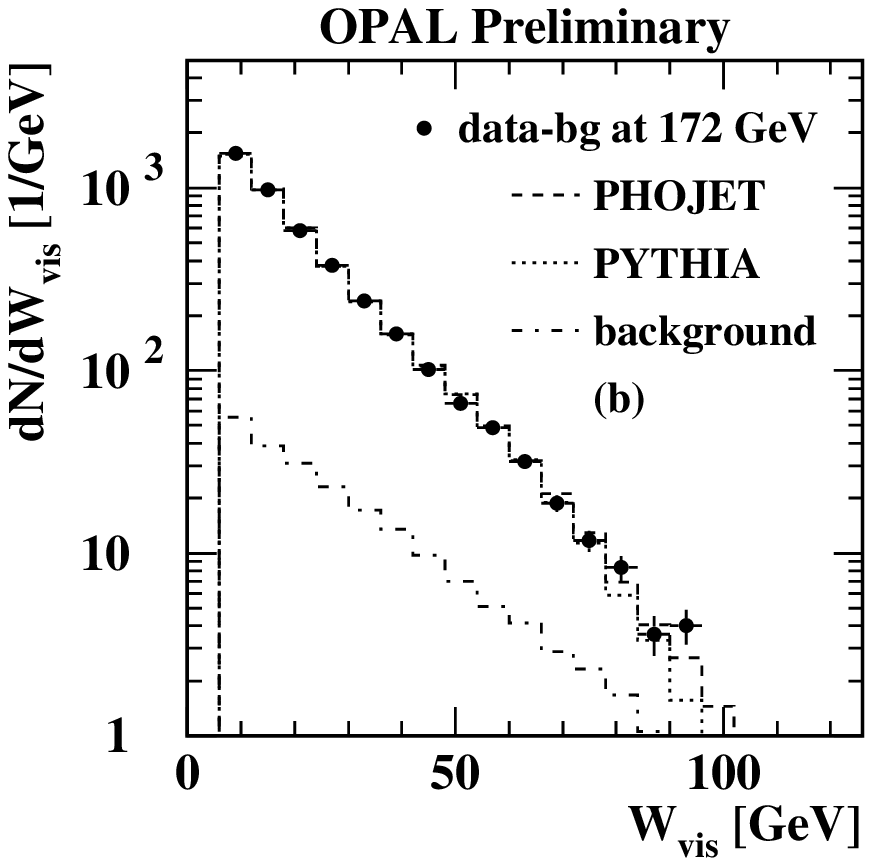}
           }
   \end{center}
\vspace*{-4em}
\caption{$\Wvis$ distribution
for all selected events with $W_{\rm vis}>6$~GeV
after background (bg) subtraction 
at 
$\sqee= 172$~GeV compared to 
MC predictions.
Statistical errors only are shown. 
}
\label{fig-wvis}
\end{figure}

A cut $W_{\rm vis}>6$~GeV is applied to all preselected events.
The $\Wvis$ distribution ${\rm d}N/{\rm d}\Wvis$ 
measured at 
$172$~GeV is shown in Fig.~\ref{fig-wvis}
where $N$ is the number of selected events.
They are well described by the MC simulations which
have been normalized to the number of data events.

\section{Unfolding of the hadronic cross section}
\begin{figure}[htb]
   \begin{center}
\mbox{
  \epsfxsize=0.465\textwidth
  \epsffile{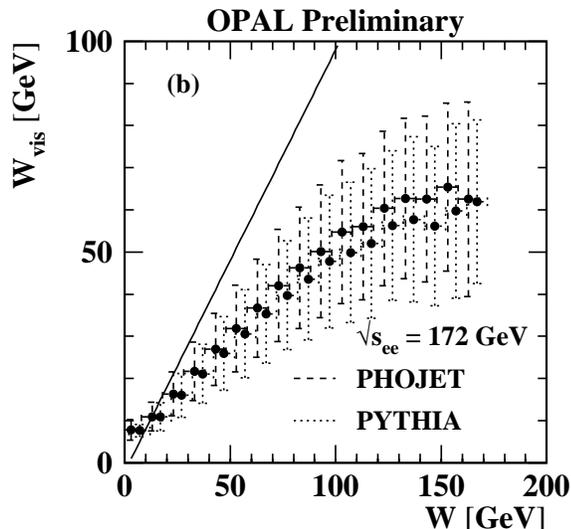} 
}
   \end{center}
\vspace*{-4em}
\caption{Correlation between $\Wvis$
and the generated invariant mass $W$ 
at $\sqee= 172$~GeV
for MC events. 
The vertical bars show the standard deviation (spread) of
the $W_{\rm vis}$ distribution in each bin.
For illustration purposes the points have been shifted by $\pm 2$~GeV.}
\label{fig-corr}
\end{figure}

\begin{figure}[htb]
   \begin{center}
\mbox{
\epsfxsize=0.465\textwidth
\epsffile{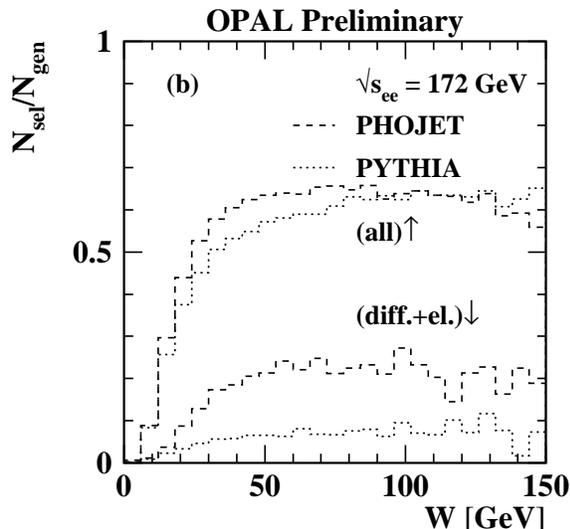} 
}
   \end{center}
\vspace*{-4em}
\caption{
The ratio of the number of selected events, $N_{\rm sel}$, to the number
of generated events, $N_{\rm gen}$, at a given generated 
invariant mass $W$ at 
$\sqee= 172$~GeV
for MC events.
The lower curves give this ratio for the diffractive and elastic
events separately.
}
\label{fig-acc}
\end{figure}

The differential cross-section $\dW$ for
the process $\ee\rightarrow(\ee+\mbox{~hadrons})$ has
to be obtained from the $W_{\rm vis}$ distribution. 
The correlation between $\Wvis$
and the generated invariant mass $W$ 
for all selected PHOJET and PYTHIA events
is shown in Fig.~\ref{fig-corr}. 
The correlation is not very good due to
hadrons which are emitted at small polar angles $\theta$.
These hadrons
are either lost in the beam pipe or they
are only detected with low efficiency in the electromagnetic
calorimeters in the forward regions (FD and SW).
The acceptance for PYTHIA
is about 15\% lower than for PHOJET at $W=40$~GeV and it approaches
the PHOJET acceptance of 60--65\% for $W>80$~GeV (Fig.~\ref{fig-acc}).
The unfolding of these resolution effects, the
correction for the detector acceptance and
the background subtraction is 
done by applying the unfolding program RUN~\cite{bib-blobel}.
The subtracted background does not include the remaining
beam-gas and beam-wall interactions.
Since the chosen bin size is not much larger
than the resolution, bin-to-bin correlations are still sizeable.

The differential cross-section $\dsee$ of the process 
$\ee\rightarrow(\ee+\mbox{~hadrons})$ can be translated into
the cross-section $\sigmagg$ for the process \ggh{} using
the luminosity function $L_{\gg}$ for the photon 
flux~\cite{bib-phojet},\cite{bib-budnev},\cite{bib-schuler}
$$
\frac{\dsee}{{\rm d} y_1 {\rm d} Q_{1}^{2}{\rm d} y_2 {\rm d} 
Q_{2}^{2}}\!=\!\sigmagg(W,Q_1^2,Q_2^2)\!\frac{{\rm d}^4 L_{\gg}}{
{\rm d} y_1 {\rm d} Q_{1}^{2}{\rm d} y_2 {\rm d} Q_{2}^{2}} ,
$$
where $y_1$ and $y_2$ denote the fraction of the beam energy
carried by photons with $y_1y_2\approx W^2/s_{\rm ee}$ (neglecting $Q^2$).
The cross-section for real photons ($Q^2=0$) is derived by
using appropriate form factors $F(Q^2)$ which describe
the $Q^2$ dependence of the hadronic cross-section:
\begin{displaymath}
\sigmagg(W,Q_1^2,Q_2^2) = F(Q^2_1)F(Q^2_2)\sigmagg(W,0,0)
\end{displaymath}
The luminosity function \Lgg{} and the form factors $F(Q^2)$
for the various \Wgg{} bins are obtained
by applying the program PHOLUM~\cite{bib-phojet}.
PHOLUM takes into account both transverse
and longitudinally polarized photons. 
The uncertainty of the extrapolation to $Q^2=0$ is estimated
to be 5--7.5\% by comparing the GVDM model 
with a simple $\rho^0$ form factor~\cite{bib-phojet}.
This uncertainty is not included in the systematic error of the
measurement.

\section{Systematic errors}
The two data samples at $\sqee=161$~GeV and $172$~GeV 
were independently analysed and the results
for the total hadronic two-photon cross-section $\sigmagg$
are found to be in good agreement and are therefore averaged.

Several distributions of the data are compared to the PYTHIA
and PHOJET simulations after detector simulation in order to
study whether the general description of the data by
the MC is sufficiently good to use it for the unfolding
of the cross-section. The MC distributions are
all normalized to the data luminosity and the background 
including the e$\gamma$ events with $Q^2>1$~GeV$^2$ is
subtracted from the data. 

In both MC models about 20\% of the cross-section is
due to diffractive and elastic events (e.g.~$\gg\rightarrow\rho\rho$).
This fraction is almost independent of $W$ for $W>10$~GeV.
The selection efficiency for the diffractive and elastic
events is very small and, although the rate is almost the
same in both models, the selection efficiencies are very
different. For a generated $W$ of 50~GeV only about 6\%
of all generated diffractive and elastic events are selected
in PYTHIA, whereas about 20\% are selected in PHOJET (Fig.~\ref{fig-acc}).
Due to the small acceptance the 
detector correction has to rely heavily on the MC simulation
for this class of events. 

\begin{figure}[htb]
   \begin{center}
\mbox{
\epsfxsize=0.465\textwidth
\epsffile{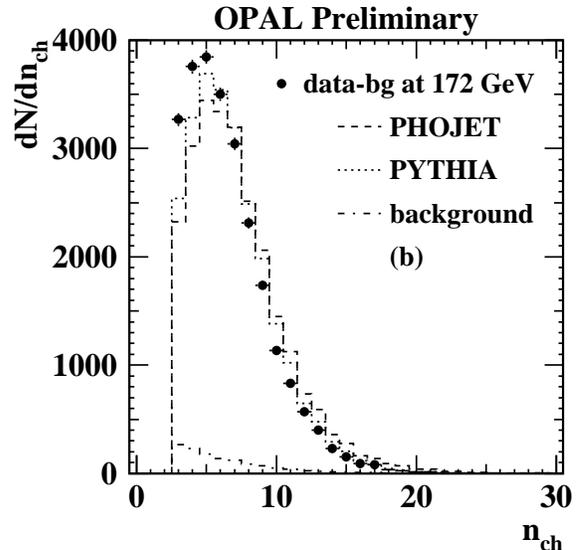} 
}
   \end{center}
\vspace*{-4em}
\caption{
$n_{\rm ch}$ distribution
for all selected events with $W_{\rm vis}>6$~GeV
after background subtraction 
at 
$\sqee= 172$~GeV 
compared to 
MC predictions.
Statistical errors only are shown.
}
\label{fig-nch}
\end{figure}

Significant discrepancies are found in the distribution
of the charged multiplicity $n_{\rm ch}$ measured in
the tracking chambers (Fig.~\ref{fig-nch}).
Both MC models significantly underestimate
the fraction of low-multiplicity events with $n_{\rm ch}<6$ 
and overestimate the fraction of high-multiplicity events in comparison
to the data.

\begin{figure}[htb]
\begin{center}
\mbox{
\epsfxsize=0.465\textwidth
\epsffile{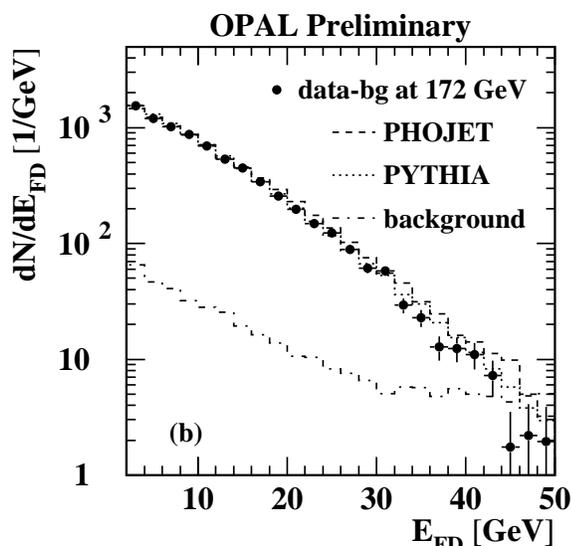} 
}
\end{center}
\vspace*{-4em}
\caption{The distribution of the energy in the forward calorimeter, 
$E_{\rm FD}$,
for all selected events with $W_{\rm vis}>6$~GeV
after background subtraction 
at 
$\sqee= 172$~GeV 
compared to 
MC predictions.
Statistical errors only are shown.
}
\label{fig-fd}
\end{figure}

The energy $E_{\rm FD}$ measured in the forward detectors (FD) is 
shown in Fig.~\ref{fig-fd} for all selected events with
$E_{\rm FD}>2$~GeV. The good agreement of data and MC
at large $E_{\rm FD}$ shows that
the background from multihadronic Z$^0$ events
and deep-inelastic e$\gamma$ events is small and that
this remaining background is reasonably well described by the MC

\begin{figure}[htb]
\begin{center}
\mbox{
\epsfxsize=0.465\textwidth
\epsffile{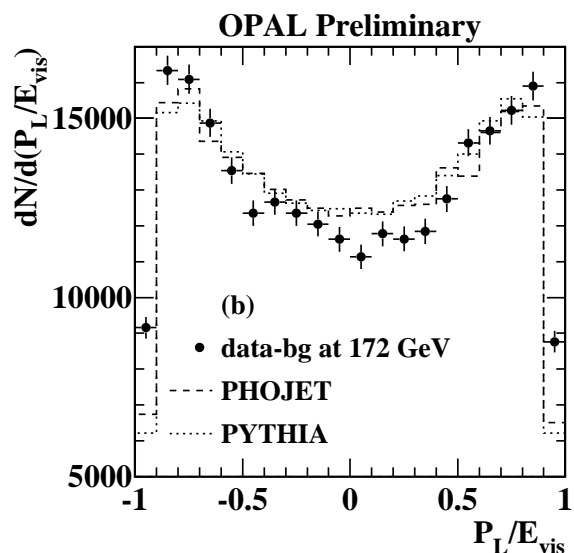} 
}
\end{center}
\vspace*{-4em}
\caption{The distribution 
of the ratio $P_{\rm L}/E_{\rm vis}$ 
for all selected events with $W_{\rm vis}>6$~GeV
after background subtraction 
at $\sqee= 172$~GeV compared to 
MC predictions.
Statistical errors only are shown.
}
\label{fig-pl}
\end{figure}

Finally we plot the ratio $P_{\rm L}/E_{\rm vis}$ 
of the longitudinal
momentum vector $P_{\rm L}$ 
to the visible total energy $E_{\rm vis}$
(Fig.~\ref{fig-pl}). 
The ratio $P_{\rm L}/E_{\rm vis}$ is
peaked around 0.9 due to the Lorentz boost of the hadronic
system. Data and MC are in reasonable agreement, different
from the observation in Ref.~\cite{bib-l3}. 
Studies of beam-gas and beam-wall events show that most of these events
have $P_{\rm L}/E_{\rm vis}>0.85$. The excess of the data
over the MC seen at large $P_{\rm L}/E_{\rm vis}$ is
therefore consistent with 2\% remaining background from
beam-gas beam-wall events.

Based on these observations
the following systematic errors are taken into account
in the measurement of the total cross-section:
\begin{itemize}
\item
Both MC models describe the data equally well.
We therefore average the results of the unfolding.
The difference between this cross-section 
and the results obtained by using PYTHIA and PHOJET alone
are taken as systematic error.
\item
For data and MC the cut 
on the charged multiplicity $n_{\rm ch}$ was increased from 
$n_{\rm ch}\ge 3$ to $n_{\rm ch}\ge 5$. This  
systematically shifts the cross-sections to lower values, mainly
at small $W$. 
The unfolding was also repeated using
only data and MC events with $3\le n_{\rm ch} \le 9$.
This variation systematically increases the cross-section 
especially at high $W$ where the average charged
multiplicity is higher than at low $W$. 
The shifts are used as systematic errors.
\item
The systematic error due to the uncertainty in  the energy scale 
of the ECAL was estimated by varying the reconstructed
ECAL energy in the MC by $\pm 5$\%.
\item
An overall normalisation uncertainty of 1\% is due to
the error on the luminosity measurement.
\item
The lower limit on the
trigger efficiency is taken into account by an additional systematic error of 
3\% in the range $10<W<38$~GeV.
\item 
It is estimated that about 2\% of the selected events could
be due to beam-gas or beam-wall interactions. This value
is therefore taken as additional systematic error.
\end{itemize}
For the total error the statistical 
and the systematic errors are added in quadrature. 
The values are given in Table~\ref{tab-ggxs}.
\begin{table*}[htbp]
  \begin{center}
\begin{tabular}{|c||c|c|c|c|c|c|}
\hline
$W$ range (GeV) &
10 --\pz16 & 16 --\pz26 & 26 --\pz38 & 38 --\pz56 & 56 --\pz80 & 80 --110    \\
\hline\hline
$\sigmagg$ [nb]&     
$385$        & $394$    & $398$      & $433$      & $485$      & $536$       \\
\hline
stat.error &    
$\pm\pz4$     & $\pm\pz3$ & $\pm\pz4$ & $\pm\pz6$ & $\pm\pz8$ & $\pm15$\\
\hline
MC model &    
$\pm26$      & $\pm\pz4$ & $\pm20$ & $\pm21$ & $\pm21$ & $\pm43$\\
$n_{\rm ch}$ cut &     
$^{+\pzz5}_{-\pz42}$ &  
$^{+\pz13}_{-\pz37}$ &  
$^{+\pz24}_{-\pz25}$ &  
$^{+\pz45}_{-\pz21}$ &  
$^{+\pz74}_{-\pz27}$ &  
$^{+\pz95}_{-\pz34}$ \\
ECAL scale &    
$\pm37$      & $\pm28$   & $\pm13$ & $\pm\pz6$ & $\pm\pz6$ & $\pm\pz7$\\
luminosity &    
$\pm\pz4$   & $\pm\pz4$  & $\pm\pz4$ & $\pm\pz4$ & $\pm\pz5$ & $\pm\pz5$\\
trigger &    
$+12$     & $+12$    & $+12$ &            &            &           \\
beam-gas &    
$-\pz8$   & $-\pz8$  & $-\pz8$ & $-\pz9$ & $-10$ & $-11$\\ \hline
total syst. &
$^{+\pz47}_{-\pz62}$ &  
$^{+\pz34}_{-\pz48}$ &  
$^{+\pz37}_{-\pz36}$ &  
$^{+\pz50}_{-\pz32}$ &  
$^{+\pz78}_{-\pz36}$ &  
$^{+105}_{-\pz56}$ \\\hline 
total error &    
$^{+\pz47}_{-\pz62}$ &  
$^{+\pz34}_{-\pz48}$ &  
$^{+\pz37}_{-\pz36}$ &  
$^{+\pz50}_{-\pz32}$ &  
$^{+\pz78}_{-\pz37}$ &  
$^{+106}_{-\pz58}$ \\
\hline
    \end{tabular}
    \caption{The total hadronic two-photon cross-section $\sigmagg$
    and the contributions from the various systematic errors.}
    \label{tab-ggxs}
  \end{center}
\end{table*}

\section{Results}
\label{sec-cross}
\begin{figure}[htb]
   \begin{center}
      \mbox{
          \epsfxsize=0.51\textwidth
          \epsffile{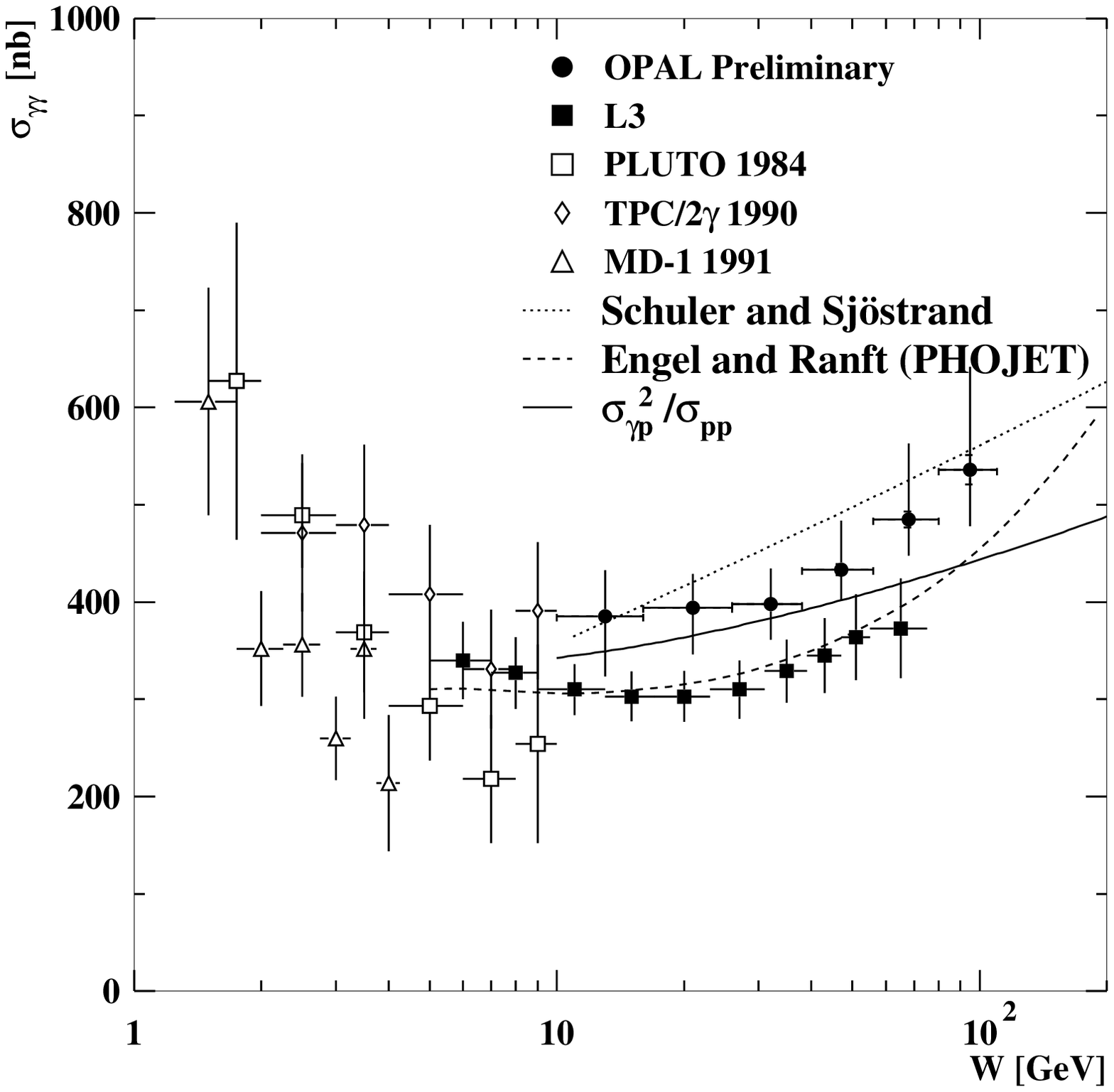}
           }
   \end{center}
\vspace*{-4em}
\caption{
$\sigmagg$ as a function of $W$.
The OPAL measurement
is compared to the measurements by
PLUTO~\cite{bib-pluto}, TPC/2$\gamma$~\cite{bib-tpc},
MD1~\cite{bib-md1} and L3~\cite{bib-l3}. The inner error
bars, which are sometimes smaller than the symbol size,
give the statistical errors and the outer error bars 
the total errors.
The data are compared to model predictions based 
on a Donnachie-Landshoff fit to total cross-sections~\cite{bib-DL}.
The solid line gives the prediction using
equation \ref{eq-tot2}.
The dotted line is the model of Schuler and
Sj\"ostrand~\cite{bib-GSTSZP73}. The model of Engel and 
Ranft~\cite{bib-phojet} used in PHOJET is shown
as dashed line.
}
\label{fig-ggxs}
\end{figure}

The total hadronic cross-section $\sigmagg$ for the process 
$\gg\rightarrow\mbox{~hadrons}$ is shown in Fig.~\ref{fig-ggxs}. 
%
In the region $W\approx 10$~GeV the OPAL measurement is 
in agreement with the measurements
at lower energies by PLUTO~\cite{bib-pluto}, TPC/2$\gamma$~\cite{bib-tpc}
within the experimental errors. 

The OPAL measurement shows
the rise in the $W$ range $10<W<110$~GeV which
is characteristic for hadronic cross-sections.
A similar rise was observed by the L3 experiment~\cite{bib-l3},
but the values for $\sigmagg$ are about 20\% lower
than the OPAL measurements. Several things should
be noted which can explain parts of this discrepancy:
First, the errors are strongly correlated between the $W$ bins
in both experiments. Secondly, L3 has used PHOJET for
the unfolding, whereas for the OPAL measurement 
the unfolding results of PHOJET and PYTHIA are
averaged. The unfolded cross-section using PHOJET
is about 5\% lower than the central value.
In both experiments the cross-sections obtained
using PHOJET are lower than the cross-section obtained
with PYTHIA. 

The total cross-section $\sigmagg$ is compared to several theoretical models.
Based on the Donnachie-Landshoff model~\cite{bib-DL}, 
we test the assumption of a universal high energy behaviour of 
$\gg$, $\gamma$p  and pp cross-sections.
The total cross-sections $\sigma$ for hadron-hadron and $\gamma$p 
collisions are well described by a Regge parametrisation of the form
\begin{equation}
\sigma=X s^{\epsilon}+Y s^{-\eta}, 
\label{eq-tot1}
\end{equation}
where $\sqrt{s}$ is the centre-of-mass energy of the hadron-hadron
or $\gamma$p interaction.
The first term in the equation
is due to Pomeron exchange and the second term
is due to Reggeon exchange~\cite{bib-pdg}.
The factors
$\epsilon = 0.0790 \pm 0.0011$ and $\eta = 0.4678 \pm 0.0059$ 
are assumed to be universal 
and have been taken from Ref.~\cite{bib-pdg}
together with the process dependent fit values of the parameters $X$ and $Y$
for the total hadronic $\gamma$p and pp cross-sections.
Assuming factorisation of the 
Pomeron term $X$, the total $\gg$ cross-section can be related
to the pp (or $\ppbar$) and $\gamma$p total cross-sections at 
high centre-of-mass energies 
$\sqrt{s}_{\gg}=\sqrt{s}_{\rm \gamma p}=\sqrt{s}_{\rm pp }$ where the Pomeron trajectory
should dominate by
\begin{equation}
\sigma_{\gg}
=\frac{\sigma_{\gamma{\rm p}}^2}{\sigma_{\rm pp }}.
\label{eq-tot2}
\end{equation}
This simple ansatz gives a reasonable
description of $\sigmagg$.
Schuler and Sj\"ostrand \cite{bib-GSTSZP73} give a 
total cross-section for the sum of all possible event
classes in their model of $\gg$ scattering where the photon
has a direct, an anomalous and a VMD component.
They consider the spread between this prediction and
the simple factorisation ansatz as conservative estimate
of the theoretical band of uncertainty.
We also plot the prediction of Engel and Ranft~\cite{bib-phojet}
which is implemented in PHOJET. It is in good agreement with
the L3 measurement and significantly lower than the OPAL
measurement. 
The steeper rise predicted by Engel and Ranft is in agreement
with both measurements.

\section{Conclusions}
\label{sec-conclusions}
We have measured the total cross-section of the process
$\gg\rightarrow\mbox{~hadrons}$ 
in the range $10< W < 110$~GeV
using the OPAL detector at LEP. 

Both MC models used fail to describe
several distributions related to the hadronic final state
like the charged multiplicity distribution. 
Further improvements of the description of the hadronic
final state are necessary to reduce the systematic error
of the measurement. It will also be important to 
gain a better understanding of the diffractive and
elastic processes for which the detection efficiency is found to be small.

With the LEP2 data the high energy behaviour of the total 
$\gg$ cross-section can be studied for the first time,
extending the accessible $W$ values 
by one order of magnitude up to $W=110$~GeV.
We observe the rise of the total $\gg$  cross-section
characteristic for the high energy behaviour
of total hadronic cross-sections. A simple
model based on Regge factorisation and
a universal Donnachie-Landshoff fit to the total cross-sections
of $\gamma$p and pp data describes the data reasonably well.

\section*{Acknowledgements}
I want to thank R.~Engel for many useful discussions and for
providing the program PHOLUM
and the organizers for the nice and interesting conference.


\end{document}